\documentclass[12pt,draftclsnofoot,onecolumn]{IEEEtran} 
\ifCLASSINFOpdf
\else
\fi

\usepackage{graphicx,color}				
\usepackage{amssymb,amsmath,amsfonts,amsthm}  

\usepackage{epstopdf}
\usepackage{lipsum} 

\usepackage{algorithm,caption}
\usepackage[noend]{algpseudocode}

\begin{document}
%
\title{Ricean K-factor Estimation based on Channel Quality Indicator in OFDM Systems using Neural Network}

\author{\IEEEauthorblockN{Kun Wang} \\
\IEEEauthorblockA{kunwang@ieee.org}
}




%


\maketitle

\begin{abstract}
Ricean channel model is widely used in wireless communications to characterize the channels with a line-of-sight path. 
The Ricean K factor, defined as the ratio of direct path and scattered paths, provides a good indication of the link quality. 
Most existing works estimate K factor based on either maximum-likelihood criterion or higher-order moments, 
and the existing works are targeted at K-factor estimation at receiver side. 
In this work, a novel approach is proposed. 
Cast as a classification problem, the estimation of K factor by neural network provides high accuracy. 
Moreover, the proposed K-factor estimation is done at transmitter side for transmit processing, thus saving the limited feedback bandwidth.
\end{abstract}


%
\IEEEpeerreviewmaketitle

\section{Introduction}  \label{sec:intro}

In wireless communications, Ricean channel model is widely adopted \cite{wang2016diversity}.
When there is a line-of-sight (LoS) path between the transmitter and the receiver,
the baseband channel can be written as the sum of a complex exponential and a complex Gaussian process, 
which are known as the ``LoS component'' and the ``diffuse component'', respectively. 
The ratio of the powers of the LoS component to the diffuse component is the Ricean factor, 
which measures the relative strength of the LoS, and hence is a measure of link quality.
Concretely, the Ricean channel coefficient can be modeled as 
\begin{equation} \label{eq:ricean_coeff}
h = \sqrt{\frac{KE}{K+1}} e^{j \phi_0} + \sqrt{\frac{E}{K+1}} \tilde{h},
\end{equation}
where $K$ is the Ricean factor, $E$ is the channel power gain, $\phi_0$ is LoS phase, and $\tilde{h}$ is the diffuse component following $\mathcal{CN}(0, 1)$. A special case is when $K=0$, it reduces to Rayleigh fading channel.

The relative power of the LoS component, represented by the $K$ factor, 
is a useful measure of the communication link quality. 
Therefore, estimation of $K$ is of practical importance in a variety of wireless scenarios, 
including channel characterization, link budget calculations, adaptive modulation, 
and geolocation applications \cite{greenwood1992mobile}, \cite{pahlavan1998wideband}.
Moreover, recent advances in space-time coding have shown that the capacity and performance of
multiple-input multiple-output (MIMO) systems depend on the Ricean factor \cite{tarokh1998space}. 
This has led \cite{catreux2002adaptive} to consider adaptive modulation schemes for MIMO systems where the adaptation
is based on the Ricean factor rather than the instantaneous channel coefficients. Hence, estimation of the Ricean factor
is important not only for channel characterization, but also in adaptive modulation schemes.

Previous works on estimation of the Ricean $K$ factor are mainly in two categories: 
maximum likelihood (ML) estimation and moment-based estimation.
In \cite{greenstein1999moment} and \cite{abdi2001estimation}, several moment-based estimators for the Ricean $K$ factor were proposed. 
These estimators use independent and noiseless channel samples. 
In \cite{tepedelenlioglu2003ricean} and \cite{azemi2004ricean}, the authors derived moment-based estimators for the Ricean $K$ parameter using
correlated and noiseless channel samples. 
The estimator proposed in \cite{azemi2004ricean} is also robust to shadowing. 
Later, in \cite{chen2005estimators}, ML and moment-based estimators for the Ricean $K$ factor
were developed using independent and noisy channel samples.

However, all of the aforementioned works estimate $K$ factor at the receiver side. 
For applications like precoder design \cite{wu2015cooperative} and adaptive modulation, it would be very helpful to have $K$ value at the transmitter side.
Of course, we can transmit the estimated $K$ value through feedback channel from receiver to transmitter.
The state-of-art LTE standard, however, does not specify the feedback of $K$ factor. 
Moreover, such feedback would consume the limited feedback bandwidth. 
In this work, instead of adding extra feedback, we will estimate $K$ value at the transmitter 
based on existing feedback quantities in LTE standard -- channel quality indicator (CQI), 
which is a quantized value of received signal-to-noise ratio (SNR).
By collecting enough observations of CQI, we can do offline training. 
In particular, neural network is chosen to capture the non-linearity inherent in the data.
Note that, the $K$ factor is continuous-valued, but discrete-valued $K$ (with rounding) is often adequate for practical applications.
Thus, we do multi-class classification here instead of regression.

\section{OFDM System Model}  \label{sec:sys_model}

In an LTE communication system with multi-path channel, orthogonal frequency division multiplexing (OFDM)
is used to combat inter-symbol interference by converting the dispersive channel to flat channels on each subcarrier \cite{wang2015joint}.
The single-input single-output channel in time domain
 with delay spread $L$ is given by $\mathbf{h} = [h_0, h_1, \ldots, h_{L-1}]^T$. 
Among them, some of the $h_l$'s may be zeros.
For the non-zero taps, for example, 
the $l$-th tap $h_l = P_l e^{j \phi_l} + \mathcal{CN}(0, 2 \sigma_l^2)$, 
i.e., $h_l \sim \mathcal{CN}(P_l e^{j \phi_l}, 2 \sigma_l^2)$, whose power gain $E_l = P_l^2 + 2 \sigma_l^2$.
Note that the $K$ factor in Eq.~(\ref{eq:ricean_coeff}) is incorporated in $P_l$ and $\sigma_l$.
By assuming the multi-path channel has unit power, we have $\sum_{l=0}^{L-1} E_l = 1$. 
Further assume same $K$ factor for all taps, i.e., $K = P_l^2 / 2\sigma_l^2, $ for $ l = 0, \ldots, L-1$.

The frequency domain channel is then given by $\mathbf{H} = [H_0, H_1, \ldots, H_{N_c-1}]^T = \mathbf{F} \mathbf{h}$, where $\mathbf{F}$ is the first $L$ columns of DFT matrix and $N_c$ is the total number of subcarriers. 
Given unit transmit power, the received SNR on the $n$-th subcarrier $SNR_n = |H_n|^2 / N_o$ 
where $N_o$ is noise variance.
In LTE, the CQI is based on effective SNR that is calculated as
\begin{equation}
SNR_{eff} = \beta f^{-1} \left( \frac{1}{N_c} \sum_{n=1}^{N_c} f\left(\frac{SNR_n}{\beta}\right) \right)
\end{equation}
where we usually choose $f(x) = e^{-x}$ and $0 < \beta \leq 1$. 
Finally, the effective SNR is quantized to CQI according to the table below (first and last columns).

\begin{figure}[!htb]
\begin{center}
\includegraphics[scale=0.45]{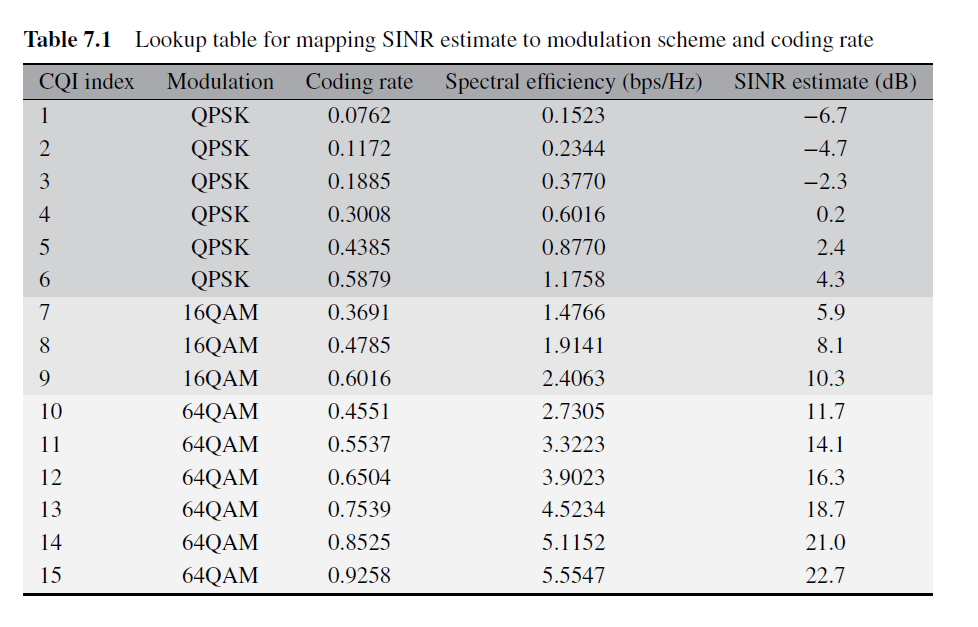}
\caption{Table for SNR to CQI mapping} \label{fig1}
\end{center}
\end{figure}

\section{Neural Network Approach}
Many experiments show that neural networks are particularly good with data that exhibit highly non-linear properties \cite{le2015tutorialpart1}.
For our case, it is easy to see that channels with lower $K$ values tend to have larger variations over time
 than those with higher $K$ values.
We use neural network in hope that such nonlinear property (variation) 
and many other properties can be captured.

\begin{figure}[!htb]
\begin{center}
\includegraphics[scale=0.45]{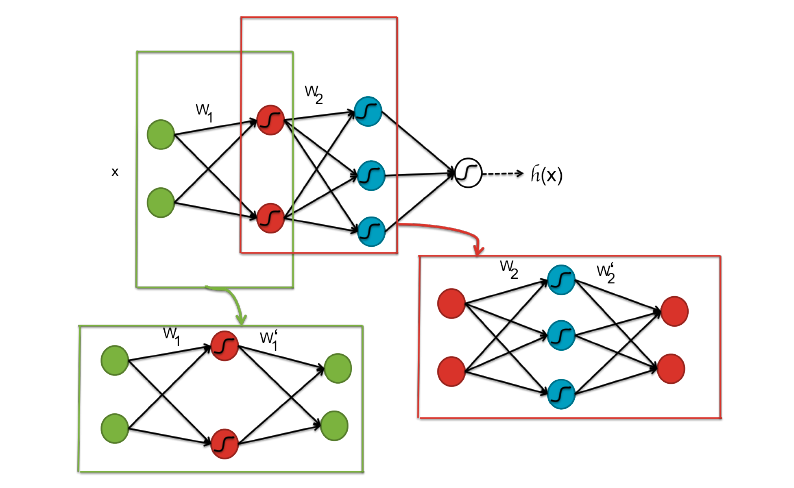}
\caption{Neural Network with Autoencoder} \label{fig2}
\end{center}
\end{figure}

Fig.~\ref{fig2} shows the structure of a neural network.
In this illustration, we have 2-dimension input (in green),
two hidden layers (in red and blue) and a final output layer. 
Generally, we let $W_{ij}^{(l)}$ denote the weight associated with the connection 
between unit $j$ in layer $l$ and unit $i$ in layer $l+1$, 
and let $\mathbf{b}^{(l)}$ denote the bias term.
Given the layer $l$'s activations $\mathbf{a}^{(l)}$, we can compute layer $l+1$'s activation $\mathbf{a}^{(l+1)}$ as
\begin{equation} \label{eq:act}
\mathbf{z}^{(l+1)} = \mathbf{W}^{(l)} \mathbf{a}^{(l)} + \mathbf{b}^{(l)}, \quad
\mathbf{a}^{(l+1)} = g(\mathbf{z}^{(l+1)}),
\end{equation}
where $g$ may be sigmoid function, or hyperbolic function, or rectified linear function.

At the final layer, the outputs should be be as close to the labels as possible.
To achieve this goal, we can set the objective of minimizing 
either mean-squared-error (MSE) or cross-entropy between the outputs and labels. 
With the objective function in hand, we can compute the gradient by using backpropagation algorithm
and thus apply gradient descent.
For the gradient descent to work, the parameters $\mathbf{W}$ and $\mathbf{b}$ need to be initialized.
It is important to ``break the symmetry'' of the neurons in the network when initialized, or in other words,
force the neurons to be different at the beginning. 
A common method is random initialization. Usually, Gaussian random or uniform random are good choices.

Another way to initialize the network is to use autoencoder.  
Autoencoders have many interesting applications, such as data compression, visualization, etc. But around
2006-2007, researchers observed that autoencoders could be used as a way to ``pretrain'' neural networks.
As a concrete example, in Fig.~\ref{fig2}, to train the red neurons, 
we will train an autoencoder that has parameters $\mathbf{W}_1$ and $\mathbf{W}'_1$.
After this, we will use $\mathbf{W}_1$ to compute the values for the red neurons for all of our data, 
which will then be used as input data to the subsequent autoencoder. 
The parameters of the decoding process $\mathbf{W}'_1$
will be discarded. The subsequent autoencoder uses the values of the red neurons as inputs, and trains an
autoencoder to predict those values by adding a decoding layer with parameters $\mathbf{W}'_2$.
In a word, the process with pretraining can be summarized as follows
\begin{itemize}
\item Pretraining step: train a sequence of shallow autoencoders, greedily one layer at a time, using
unsupervised data
\item Fine-tuning step 1: train the last layer using supervised data
\item Fine-tuning step 2: use backpropagation to fine-tune the entire network.
\end{itemize}
More details about neural networks and autoencoders can be found in \cite{le2015tutorialpart1} and \cite{le2015tutorialpart2}.

\section{Numerical Results}

\noindent
\textbf{OFDM Settings}: Bandwidth 10 MHz. Channel model is ``Pedestrian B'' model 
whose channel delay spread (in $\mu$s) is [0.0  0.2  0.8  1.2  2.3  3.7] and 
channel tap gain (in dB) is [0.0 -0.9 -4.9 -8.0 -7.8 -23.9].
The $\beta$ used in computing effective SNR is set to 0.5.

\vspace{2mm}
\noindent
\textbf{CQI Data Generation}:
CQI is selected at the receiver side based on instantaneous SNR 
which depends on channel power gain and average transmit SNR.
For each OFDM symbol, we obtain a CQI indicating the link quality. 
By accumulating over $N$ time epochs, we can have a vector of CQIs of dimension $N$. 
Since CQI is SNR-dependent and we want the classifier to work under any SNR,
the CQI data are generated at average SNR = 1:1:25 dB 
for every $K$ value, where we set $K$ = 0:1:10. 
Hence, there is a total of 275 combinations of SNR and $K$ value.
Furthermore, under each pair of SNR and $K$ value, we generate 10 CQI vectors (each of
dimension $N$) to make the classifier insensitive to channel randomness.
Finally, these data are randomly divided into training data (80\%) and test data (20\%).

\vspace{2mm}
\noindent
\textbf{Neural Network Setup}:
In this project, a neural network with one hidden layer is used. 
The input dimension is $N$ (the dimension of a CQI vector).
We set the number of neurons in the hidden layer to be 100. 
In the final output layer, there are 11 activations, 
corresponding to the 11 possible $K$ values (from 0 to 10 with step 1).
It should be pointed out that the data labels use one-hot encoding, 
meaning that $K=i$ is encoded by only a 1 on the $(i+1)$-th position and 0s elsewhere.

\vspace{2mm}
\noindent
\textbf{Algorithm and Stopping Criterion}:
To train the neural network, we choose the cross-entropy criterion and use the backpropagation algorithm
implemented in MATLAB. As a note on terminology, the term ``backpropagation" is sometimes used to refer specifically to the gradient descent algorithm, when applied to neural network training. However, that terminology is not used in MATLAB documentations, since the process of computing the gradient and Jacobian by performing calculations backward through the network is applied in many other training functions: BFGS Quasi-Newton, Scaled Conjugate Gradient (SCG), etc. 
For this project, SCG is chosen to train the network.
To terminate the training process, several stopping criteria exist: 
reaching maximum number of iterations, reaching (minimum) threshold of objective function, 
reaching (minimum) threshold of gradient value, or failing to pass validation checks.

\begin{figure}[!tb]
\begin{center}
\includegraphics[scale=0.45]{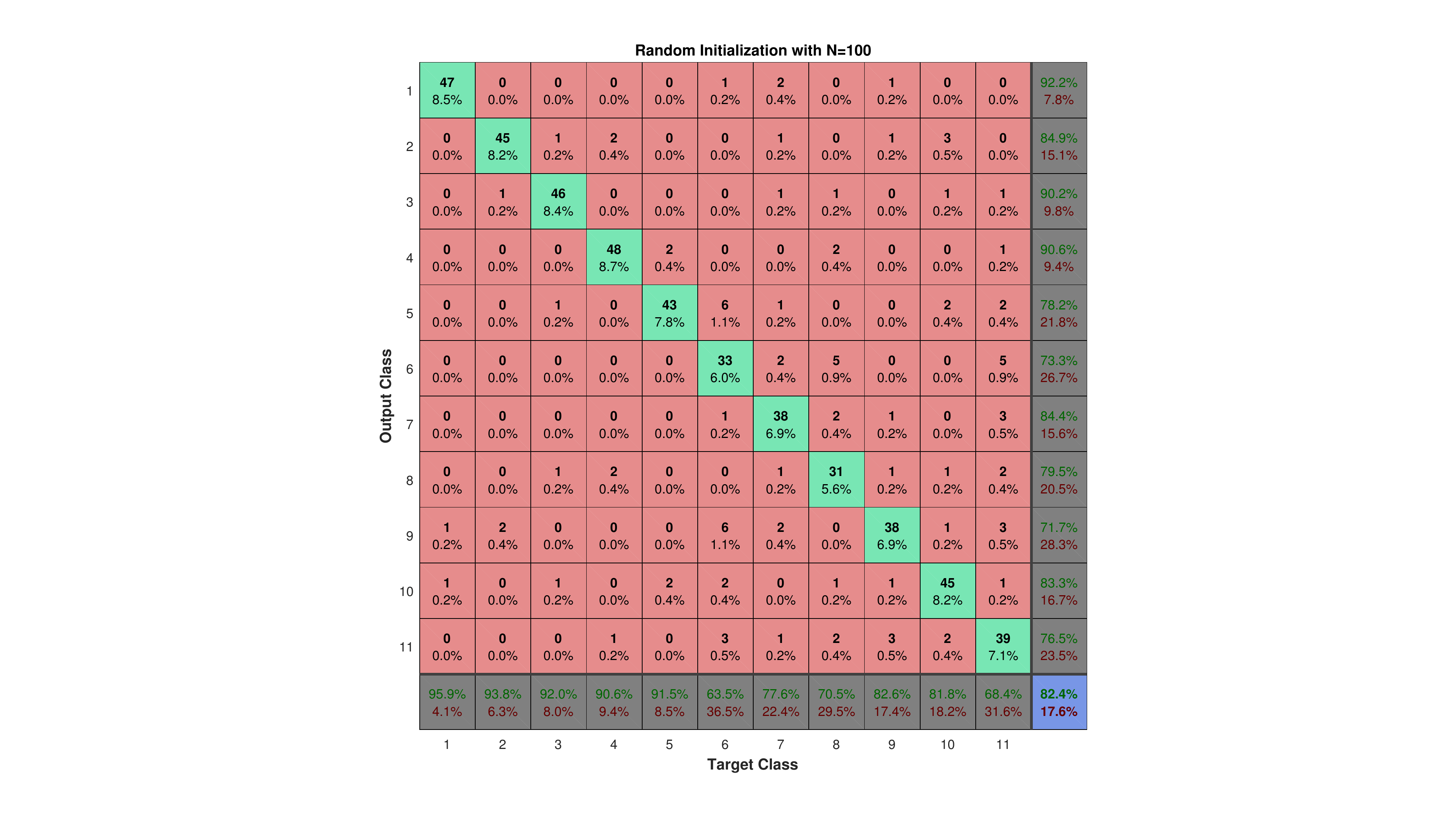}
\caption{Random Initialization with Input Dimension 100. } \label{fig3}
\end{center}
\end{figure}

\begin{figure}[!tb]
\begin{center}
\includegraphics[scale=0.45]{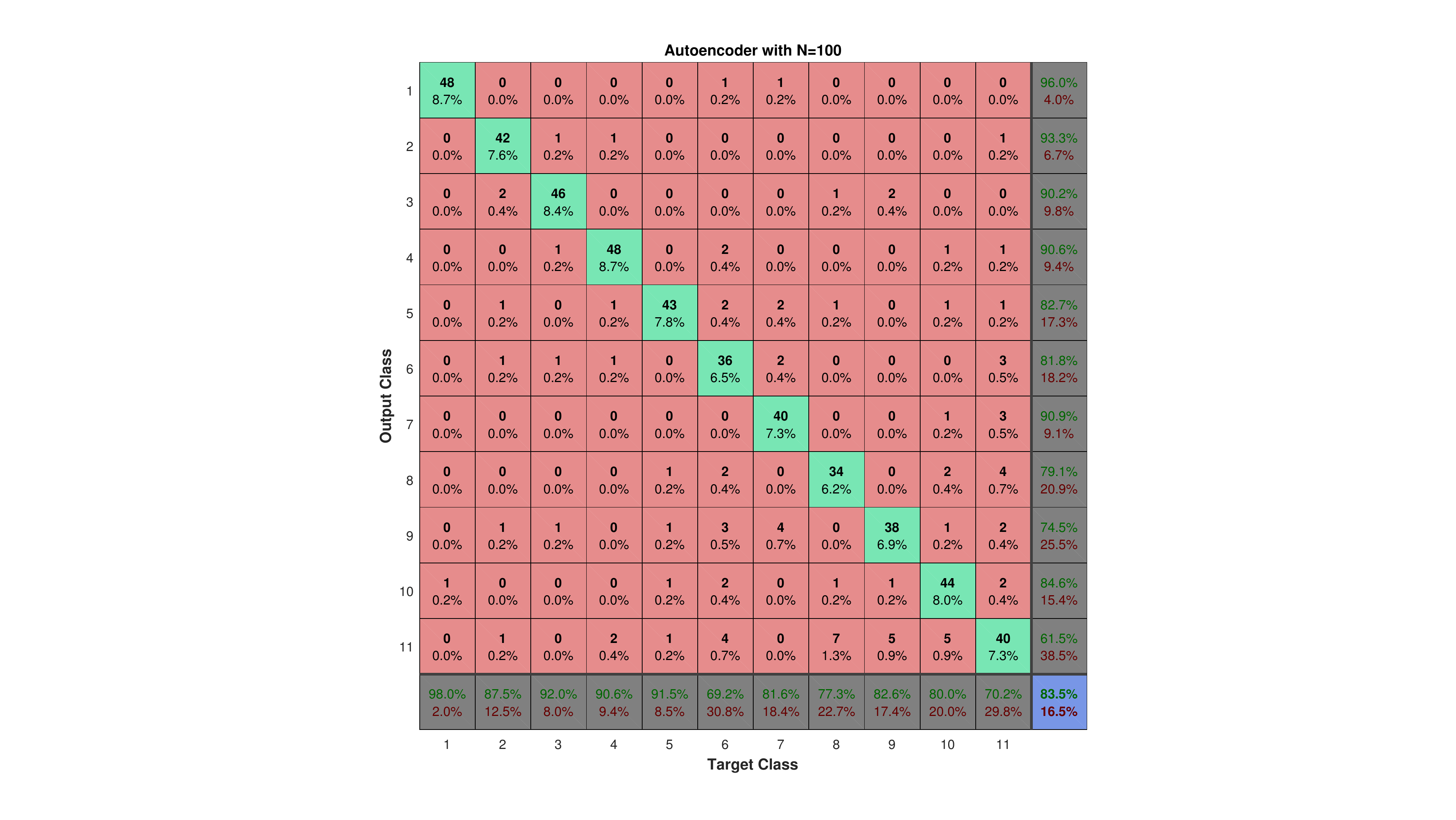}
\caption{Autoencoder  Initialization with Input Dimension 100. } \label{fig4}
\end{center}
\end{figure}

\begin{figure}[!tb]
\begin{center}
\includegraphics[scale=0.45]{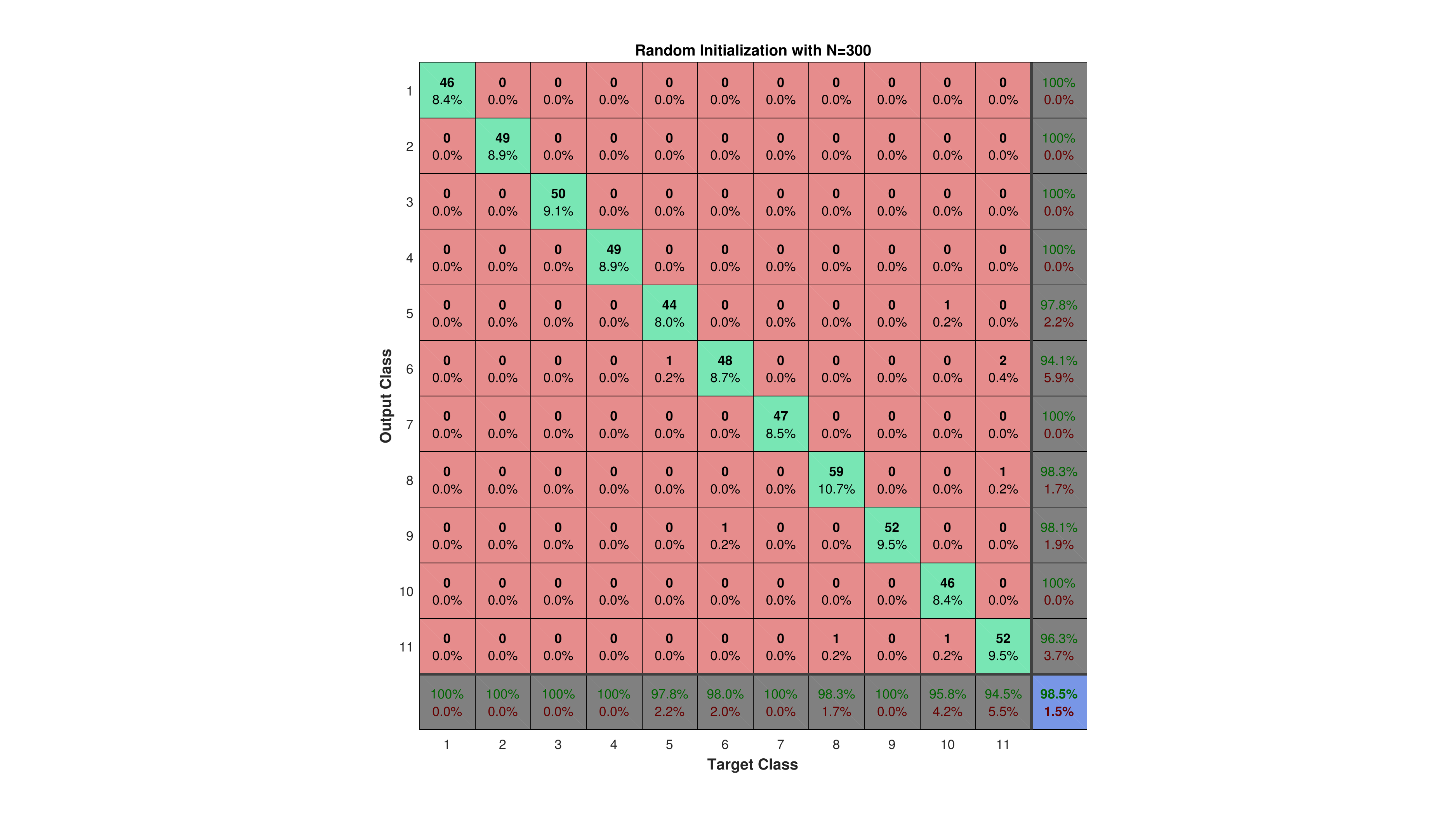}
\caption{Random Initialization with Input Dimension 300. } \label{fig5}
\end{center}
\end{figure}

\begin{figure}[!tb]
\begin{center}
\includegraphics[scale=0.45]{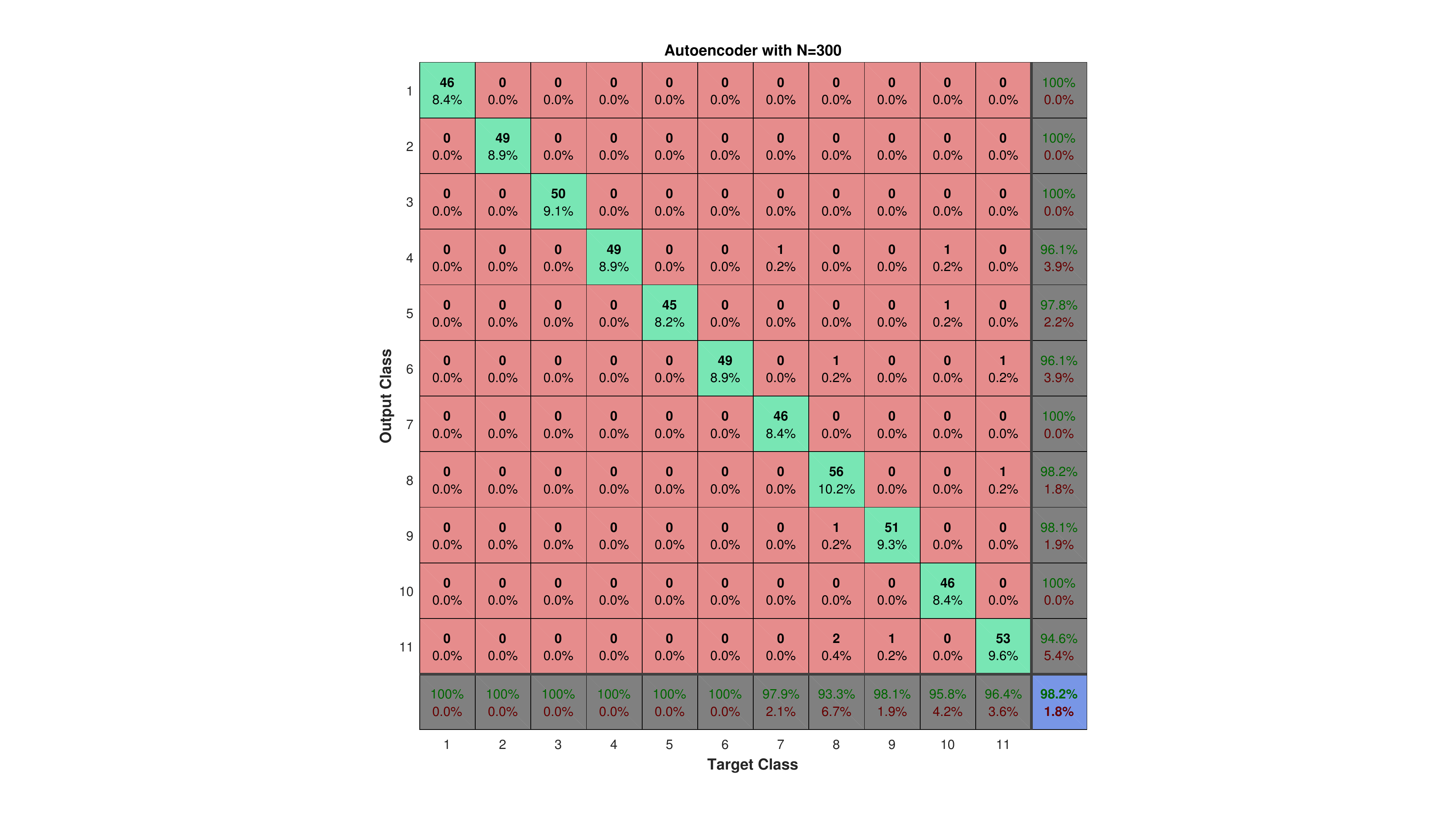}
\caption{Autoencoder Initialization with Input Dimension 300. } \label{fig6}
\end{center}
\end{figure}

\begin{figure}[!tb]
\begin{center}
\includegraphics[scale=0.45]{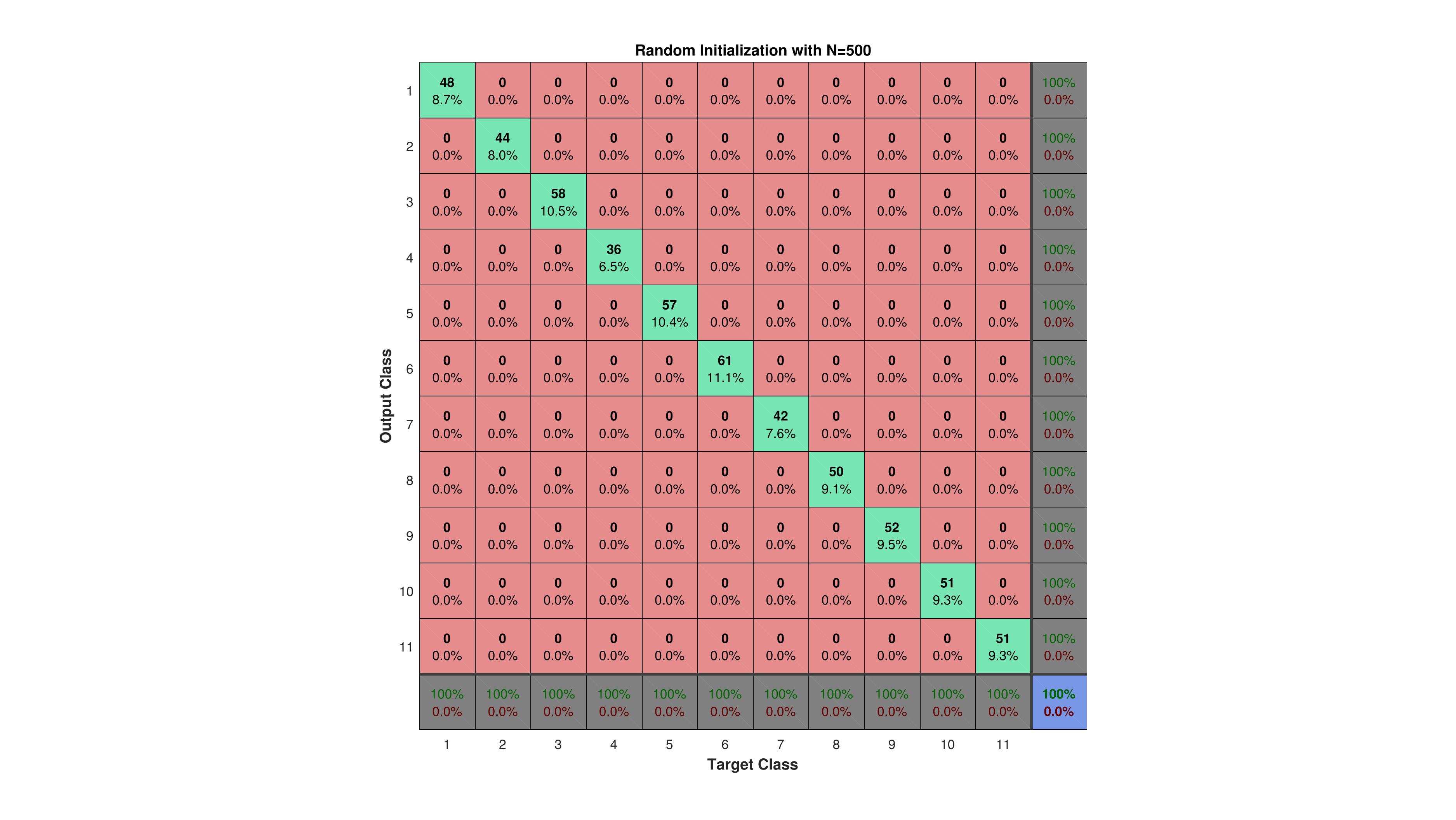}
\caption{Random Initialization with Input Dimension 500. } \label{fig7}
\end{center}
\end{figure}

\begin{figure}[!tb]
\begin{center}
\includegraphics[scale=0.45]{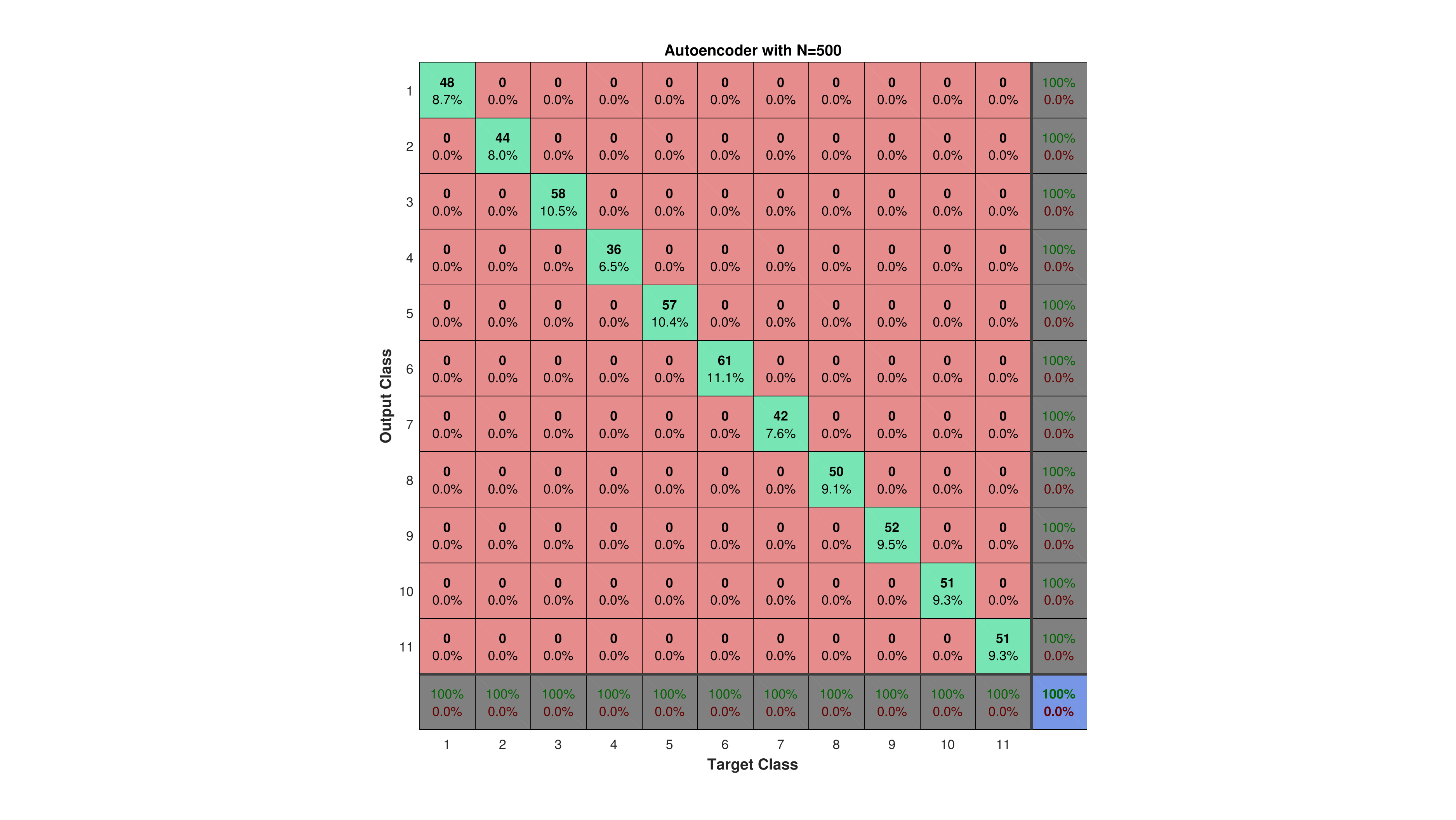}
\caption{Autoencoder Initialization with Input Dimension 500. } \label{fig8}
\end{center}
\end{figure}

\vspace{2mm}
\noindent
\textbf{Performance Evaluation}:
The classification performance is evaluated by showing the confusion matrix in Fig.~\ref{fig3} - Fig.~\ref{fig8}. 
Neural networks with both random initialization and autoencoder initialization are tested. 
Moreover, the input dimensions are varying: $N=100, 300, 500$.
The overall correct and wrong classification rates are indicated in the lower right corner of every confusion matrix
in green and red, respectively. 
It is clear that the performance improves as the dimensions go large,
and it is 100\% accuracy when $N=500$.
However, initialization with autoencoders do not show any advantages over random initialization.
This is probably because shallow networks instead of deep networks are used here.

\section{Summary}
In this work, a classification problem is solved by using neural networks. 
Classical multi-layer perceptron neural network is used, with both random initialization 
and autoencoder initialization. 
As the input dimension increases, the classification accuracy greatly improves. 
It is interesting to see if good accuracy can be achieved with lower dimensional inputs 
when using convolutional neural networks. 
Due to hardware constraint, the experiment with convolutional neural network is left for future work.

Regarding the K-factor itself, it is also interesting to see the impact of the estimation accuracy on 
the precoder and receiver performance. 
More specifically, most of my recent works are on the topic of robust detection-decoding against 
channel uncertainty \cite{wang2018integrated,wang2014joint,wang2016robust,wang2015diversity,wang2016fec,wang2015joint,wang2016diversity,wang2017galois,wang2017phase,wang2018joint}, 
and it is worthwhile to investigate the effect of K-factor estimation errors on
receiver performance. 
Last but not the least, applying neural network to receiver design would be an interesting topic.

\bibliographystyle{IEEEtran}
\bibliography{IEEEabrv,mybibfile}
%
%
%

\end{document}